\documentclass[]{article}
\usepackage[english]{babel}
\usepackage{lineno}
\usepackage{graphicx,multicol}
\usepackage{epic,eepic,epsfig}
\usepackage{amssymb}
\usepackage{amsmath}
\usepackage{todonotes}
\usepackage{hyperref}
\newcommand{\induce}[2]{\mbox{$ #1 [#2]$}}

\newcommand{\2}{\vspace{2mm}}

\newtheorem{theorem}{Theorem}[section]
\newtheorem{lemma}[theorem]{Lemma}
\newtheorem{proposition}[theorem]{Proposition}

\newtheorem{corollary}[theorem]{Corollary}
\newtheorem{conjecture}[theorem]{Conjecture}

\newtheorem{problem}[theorem]{Problem}

\newcommand{\qed}{\hfill$\diamond$}
\newcommand{\pf}{{\bf Proof: }}

\newcommand{\emphX}[1]{\textbf{#1}}

\begin{document}
\bibliographystyle{plain}
\title{Degree-constrained $2$-partitions of graphs}

\author{J. Bang-Jensen\thanks{Department of Mathematics and Computer
    Science, University of Southern Denmark, Odense DK-5230, Denmark
    (email:jbj@imada.sdu.dk).  This work was done while the first
    author was visiting LIRMM, Universit\'e de Montpellier,
    France. Hospitality is gratefully acknowledged. The research of
    Bang-Jensen was supported by the Danish research council under
    grant number 7014-00037B.} \and St\'ephane Bessy\thanks{LIRMM,
    Universit\'e de Montpellier, France (email:
    stephane.bessy@lirmm.fr).}}

\maketitle

\begin{abstract}
A $(\delta\geq k_1,\delta\geq k_2)$-partition of a graph $G$ is a
vertex-partition $(V_1,V_2)$ of $G$ satisfying that
$\delta(G[V_i])\geq k_i$ for $i=1,2$. We determine, for all positive
integers $k_1,k_2$, the complexity of deciding whether a given graph
has a $(\delta\geq k_1,\delta\geq k_2)$-partition.\\ We also
address the problem of finding a function $g(k_1,k_2)$ such that the
$(\delta\geq k_1,\delta\geq k_2)$-partition problem is ${\cal
  NP}$-complete for the class of graphs of minimum degree less than
$g(k_1,k_2)$ and polynomial for all graphs with minimum degree at
least $g(k_1,k_2)$. We prove that $g(1,k)=k$ for $k\ge 3$, that
$g(2,2)=3$ and that $g(2,3)$, if it exists, has value 4 or 5.\\
  
  \noindent{}{\bf Keywords:} ${\cal NP}$-complete, polynomial, 2-partition, minimum degree.
\end{abstract}

\section{Introduction}

A \emphX{$2$-partition} of a graph $G$ is a partition of $V(G)$ into
two disjoint sets. Let $\mathbb{P}_1, \mathbb{P}_2$ be two graph
properties, then a \emphX{$(\mathbb{P}_1,\mathbb{P}_2)$-partition} of
a graph $G$ is a $2$-partition $(V_1,V_2)$ where $V_1$ induces a graph
with property $\mathbb{P}_1$ and $V_2$ a graph with property
$\mathbb{P}_2$. For example a $(\delta\geq k_1,\delta\geq
k_2)$-partition of a graph $G$ is a $2$-partition $(V_1,V_2)$ where
$\delta(G[V_i])\geq k_i$, for $i=1,2$

There are many papers dealing with vertex-partition problems on
(di)graphs. Examples (from a long list) are
\cite{alonCPC15,bangTCS640,bangTCS636,bazganTCS355,bazganDAM155,bensmailJCO30,bonamyMFCS2017,bonamyIPL131,chvatalJGT8,dyerDAM10,foldesCN19,gerberEJOR125,gerberAJC29,grigorievLNCS5911,hajnalC3,kanekoJGT27,kuehnJCT88,leDAM131,liuSCM58,maarXiv1706,misraJCO24,stiebitzKAM,
  stiebitzJGT23,suzukiIPL33,thomassenJGT7,vanthofTCS410,xiaoTCS659}. Examples of
2-partition problems are recognizing bipartite graphs (those having
has a $2$-partition into two independent sets) and split graphs (those
having a $2$-partition into a clique and an independent
set)~\cite{foldesCN19}. It is well known and easy to show that there
are linear algorithms for deciding whether a graph is bipartite,
respectively, a split graph. It is an easy exercise to show that every
graph $G$ has a 2-partition $(V_1,V_2)$ such that the degree of each
vertex in $G[V_i]$, $i\in [2]$ is at most half of its original
degree. Furthermore such a partition can be found efficiently by a
greedy algorithm. In \cite{gerberEJOR125,gerberAJC29} and several other papers
the opposite condition for a 2-partition was studied.  Here we require
the that each vertex has at least half of its neighbours inside the
set it belongs to in the partition. This problem, known as the
satisfactory partition problem, is ${\cal NP}$-complete for general
graphs \cite{bazganDAM154}.

A partition problem that has received particular attention is that of
finding sufficient conditions for a graph to possess a $(\delta\geq
k_1,\delta\geq k_2)$-partition. Thomassen \cite{thomassenJGT7} proved
the existence of a function $f(k_1,k_2)$ so that every graph of
minimum degree at least $f(k_1,k_2)$ has a $(\delta\geq k_1,\delta\geq
k_2)$-partition. He proved that $f(k_1,k_2)\leq
12\cdot\max\{k_1,k_2\}$. This was later improved by Hajnal
\cite{hajnalC3} and H\"aggkvist (see \cite{thomassenJGT7}). Thomassen
\cite{thomassenJGT7,thomassen1988} asked whether it would hold that
$f(k_1,k_2)=k_1+k_2+1$ which would be best possible because of the
complete graph $K_{k_1+k_2+1}$. Stiebitz \cite{stiebitzJGT23} proved
that indeed we have $f(k_1,k_2)=k_1+k_2+1$.  Since this result was published, several
groups of researchers have tried to find extra conditions on the graph
that would allow for a smaller minimum degree requirement. Among
others the following results were obtained.

\begin{theorem}\cite{kanekoJGT27}
\label{kanekothm}
For all integers $k_1,k_2\geq 1$ every triangle-free graph $G$ with
$\delta(G)\geq k_1+k_2$ has a $(\delta\geq k_1,\delta\geq
k_2)$-partition.
\end{theorem}
  
\begin{theorem}\cite{maarXiv1706}
  \label{mathm}
  For all integers $k_1,k_2\geq 2$ every graph $G$ with no 4-cycle and
  with $\delta(G)\geq k_1+k_2-1$ has a $(\delta\geq k_1,\delta\geq
  k_2)$-partition.
\end{theorem}

\begin{theorem}\cite{liuDAM226}
  \label{liuthm}
  \begin{itemize}
  \item For all integers $k_1,k_2\geq 1$, except for $K_3$, every
    graph $G$ with no $K_4-e$ and with $\delta(G)\geq k_1+k_2$ has a
    $(\delta\geq k_1,\delta\geq k_2)$-partition.
    \item For all integers $k_1,k_2\geq 2$ every triangle-free graph
      $G$ in which no two 4-cycles share an edge and with
      $\delta(G)\geq k_1+k_2-1$ has a $(\delta\geq k_1,\delta\geq
      k_2)$-partition.
    \end{itemize}
  \end{theorem}

The original proof that $f(k_1,k_2)=k_1+k_2+1$ in \cite{stiebitzJGT23}
is not constructive and neither are those of Theorems \ref{mathm} and
\ref{liuthm}. In \cite{bazganDAM155} Bazgan et al. gave a polynomial
algorithm for constructing a $(\delta\geq k_1,\delta\geq
k_2)$-partition of a graph with minimum degree at least $k_1+k_2+1$ or
at least $k_1+k_2$ when the input is triangle-free.

The main result of this paper is a full characterization of the
complexity of the $(\delta\geq k_1,\delta\geq k_2)$-partition problem. 

\begin{theorem}
\label{main}
Let $k_1,k_2\geq 1$ with $k_1\leq k_2$ be integers. When $k_1+k_2\leq
3$ it is polynomial to decide whether a graph has a 2-partition
$(V_1,V_2)$ such that $\delta(G[V_i])\geq k_i$ for $i=1,2$. For all
other values of $k_1,k_2$ it is {\cal NP}-complete to decide the
existence of such a partition.
\end{theorem}

A result in \cite{chvatalJGT8} implies that $(\delta\geq 3,\delta\geq 3)$-partition is ${\cal NP}$-complete already for 4-regular graphs and there are other results about the complexity of finding partitions with lower and/or upper bounds on the degrees inside each partition, such as \cite{bazganDAM154,bazganTCS355,gerberEJOR125,gerberAJC29,xiaoTCS659}, but we did not find anything which implies Theorem \ref{main}.

The result of Stiebitz~\cite{stiebitzJGT23} insures that if the
minimum degree of the input graph is large enough, at least $k_1+k_2+1$, then the $(\delta\geq k_1,\delta\geq k_2)$-partition
always exists. We conjecture that if this minimum degree is large but
less than $k_1+k_2+1$ then the $(\delta\geq k_1,\delta\geq
k_2)$-partition is not always trivial but can be solved in polynomial
time.

\begin{conjecture}
\label{conj:functiong}
There exists a function $g(k_1,k_2)$ so that for all $1\leq k_1\leq
k_2$ with $k_1+k_2\geq 3$ the $(\delta\geq k_1,\delta\geq
k_2)$-partition problem is ${\cal NP}$-complete for the class of
graphs of minimum degree less than $g(k_1,k_2)$ and polynomial for all
graphs with minimum degree at least $g(k_1,k_2)$.
\end{conjecture}

In the next section we introduce notions and tools that will be used
later. In Section~3 we give the proof of Theorem~\ref{main} and in
Section~4 we provide some partial results concerning
Conjecture~\ref{conj:functiong}. In particular we prove that
$g(1,k)=k$ for $k\ge 3$, that $g(2,2)=3$ and that $g(2,3)$, if it
exists, has value 4 or 5. Finally in Section~5 we address some other
partition problems mainly dealing with (edge-)connectivity in each
part of the partition.

\2 Notice that regarding the results that we establish in this paper
the first open case of Conjecture~\ref{conj:functiong} is the
following problem.
\begin{problem}
What is the complexity of the $(\delta\geq 2,\delta\geq 3)$-partition problem
for graphs of minimum degree 4?
\end{problem}

\section{Notation, definitions and preliminary results}

Notation is standard and follows \cite{bang2009,bondy2008}. In this
paper graphs have no parallel edges and no loops. We use the shorthand
notation $[k]$ for the set $\{1,2,\ldots{},k\}$.

\subsection{Special Graphs}

We first define some graphs that will be used frequently in our proofs
to ensure that certain vertices have a sufficiently high degree.

\begin{figure}[!ht]
\centering
\scalebox{0.5}{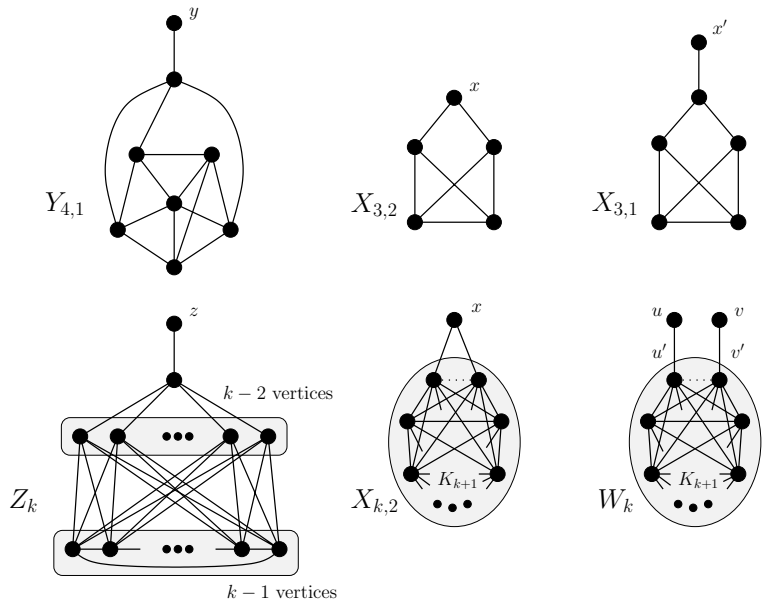}
\caption{The graphs $Y_{4,1}$, $X_{3,2}$, $X_{3,1}$, $Z_k$,
  $X_{k,2}$ and $W_k$.}
\label{fig:pendingGadgets}
\end{figure}

For all $k\geq 3$ we let $X_{k,2}$ be the graph we obtain from
$K_{k+1}$ by subdividing one edge by a vertex $x$. Let $X_{3,1}$ be
obtained from $X_{3,2}$ by adding a vertex $x'$ adjacent to the degree
2 vertex $x$ of $X_{3,2}$. Let $Y_{4,1}$ be the graph on 7 vertices
which we obtain from a 5-wheel by adding a new edge $e$ linking 2 non
adjacent vertices of the outer cycle of the 5-wheel, a new vertex
joined to the 3 vertices of this outer cycle not incident to $e$ and
to another new vertex $y$. For $k=3$ let $Z_k$ be the graph $X_{3,1}$
that we defined above and let $z=x'$. For $k\geq 4$ let $Z_k$ be the
graph that we obtain from $K_{k-2,k-1}$ by adding a cycle on the $k-1$
vertices of degree $k-2$ and then adding a new vertex $z$ adjacent to
all the $k-2$ vertices of degree $k-1$. And finally let $W_k$ be the
graph we obtain from $K_{k+1}$ by deleting one edge $u'v'$ and the
adding two new vertices $u,v$ and the edges $uu',vv'$.\\
All these graphs are depicted in Figure~\ref{fig:pendingGadgets}.

\subsection{Connected instances of 3-SAT}

For a given instance $\cal F$ of 3-SAT with clauses $C_1,C_2,\ldots
{},C_m$ and variables $x_1,\ldots{},x_n$, where each variable $x_i$
occurs at least once as the literal $x_i$ and at least once as
$\bar{x}_i$, we define the bipartite graph $B(\cal F)$ as the graph
with vertex set
$\{v_1,\bar{v_1},v_2,\bar{v}_2,\ldots{},v_n,\bar{v}_n\}\cup
\{c_1,c_2,\ldots{},c_m\}$, where the first set corresponds to the
literals and the second one to the clauses, and edge set containing an
edge between the vertex $c_j$ and each of the 3 vertices corresponding
to the literals of $C_j$ for every $j\in [m]$. We say that $\cal F$ is
a {\bf connected} instance if $B(\cal F)$ is connected.

\begin{lemma}
\label{connected3SAT}
3-SAT is ${\cal NP}$-complete for instances $\cal F$ where $B(\cal F)$ is connected
\end{lemma}

\pf Suppose $B({\cal F})$ has connected components
$X_1,X_2,\ldots{},X_k$, where $k\geq 2$. Fix a literal vertex
$\ell_i\in X_i$ for each $i\in [k]$, add a new variable $y$ and $k-1$
new clauses $C'_1,\ldots{},C'_{k-1}$ where $C'_j=(\ell_{j-1}\vee
y\vee\ell_j)$, $j\in [k-1]$. Let ${\cal F}'$ be the new formula
obtained by adding the variable $y$ and the clauses
$C'_1,\ldots{},C'_{k-1}$. It is easy to check that ${\cal F}'$ is
equivalent to $\cal F$ and that $B({\cal F}')$ is connected. \qed

\2

By adding a few extra variables, if necessary, we can also obtain an
equivalent connected instance in which each literal occurs at least
twice. We leave the easy details to the interested reader.

\subsection{Ring graphs and 3-SAT}\label{ringdefsec}

We first introduce an important class of graphs that will play a
central role in our proofs. The directed analogue of these graphs was
used in \cite{bangTCS640,bangTCS636}.  A {\bf ring graph} is the graph
that one obtains by taking two or more copies of the complete
bipartite graph on 4 vertices $\{a_1,a_2,b_1,b_2\}$ and edges
$\{a_1b_1,a_1b_2,a_2b_1,a_2b_2\}$ and joining these in a circular
manner by adding a path $P_{i,1}$ from the vertex $b_{i,1}$ to
$a_{i+1,1}$ and a path $P_{i,2}$ from $b_{i,2}$ to $a_{i+1,2}$ where
$b_{i,1}$ is the $i$th copy of $b_1$ etc and indices are 'modulo' $n$
($b_{n+1,j}=b_{1,j}$ for $j\in [2]$ etc). Our proofs are all
reductions from ${\cal NP}$-complete variants of the 3-SAT problems.  We call
the copies of $\{a_1b_1,a_1b_2,a_2b_1,a_2b_2\}$ {\bf switch vertices}.

We start by showing how we can associate a ring graph to a given 3-SAT
formula.  Let ${\cal F}= C_1\wedge C_2\wedge{}\ldots\wedge{}C_m$ be an
instance of 3-SAT consisting on $m$ clauses $C_1,\ldots{},C_m$ over
the same set of $n$ boolean variables $x_1,\ldots{},x_n$. Each clause
$C_i$ is of the form
$C_i=(\ell_{i,1}\vee{}\ell_{i,2}{}\vee{}\ell_{i,3})$ where each
$\ell_{i,j}$ belongs to
$\{x_1,x_2,\ldots{},x_n,\bar{x}_1,\bar{x}_2,\ldots{},\bar{x}_n\}$ and
$\bar{x}_i$ is the negation of variable $x_i$. By adding extra clauses
to obtain an equivalent formula, if necessary, we can ensure that
every literal occurs at least twice in $\cal F$. We shall use this
fact in one of our proofs.\\

For each variable $x_i$ the ordering of the clauses above induces an
ordering of the occurrences of $x_i$, resp $\bar{x}_i$, in the
clauses. Let $q_i$ (resp. $p_i$) denote the number of times $x_i$
(resp. $\bar{x}_i$) occurs in the clauses Let $R({\cal F})=(V,E)$ be
the ring graph defined as follows.  Its vertex set is
  $$V=\{a_{1,1},\ldots{}a_{n,1},a_{1,2}\ldots{},a_{n,2}\}\cup
  \{b_{1,1},\dots{},b_{n,1},b_{1,2},\ldots{},b_{n,2}\}\cup 
  \bigcup_{i=1}^n\{v_{i,1},\dots{},v_{i,q_i},v'_{i,1},\ldots{},v'_{i,p_i}\}$$
Its edge set $E$ consists of the following edges:
\begin{itemize}
\item $\bigcup_{i=1}^n\{a_{i,1}b_{i,1},a_{i,1}b_{i,2},a_{i,2}b_{i,1},a_{i,2}b_{i,2}\}$  
\item the edges  of the paths $P_{i,1},P_{i,2}$, $i\in [n]$ where
  $P_{i,1}=b_{i,1}v_{i,1}\ldots{}v_{i,q_i}a_{i+1,1}$ and\\
  $P_{i,2}=b_{i,2}v'_{i,1}\ldots{}v'_{i,p_i}a_{i+1,2}$.
\end{itemize}

\begin{figure}[h!]
\centering
\scalebox{0.6}{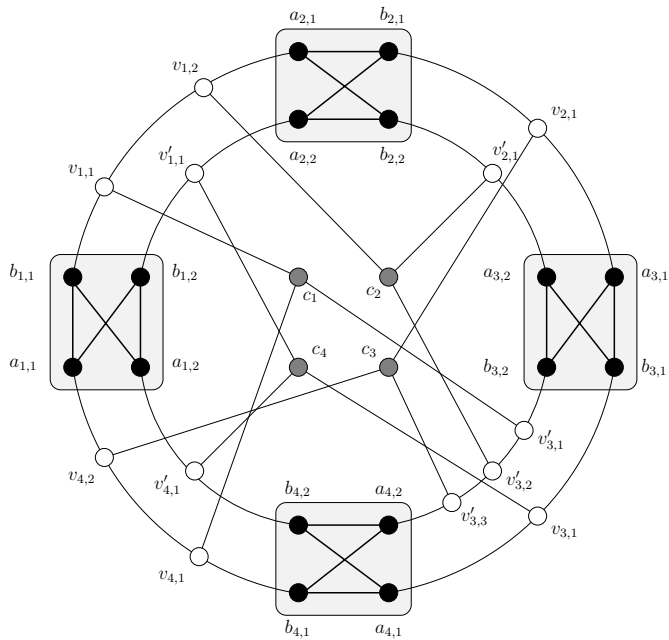}
\caption{The ring graph $R(\cal F)$ corresponding to the formula
  ${\cal
    F}=(x_1\vee{}x_4\vee{}\bar{x}_3)\land{}(x_1\vee{}\bar{x}_2\vee{}\bar{x}_3)\land{}(x_2\vee{}\bar{x}_3\vee{}x_4)\land{}(\bar{x}_1\vee{}x_3\vee{}\bar{x}_4)$. The
  grey boxes contain the switch vertices, the white vertices are the
  variable vertices and we added the clauses vertices $c_1$, $c_2$,
  $c_3$ and $c_4$ (these are not part of the ring graph $R(\cal F)$).}
\label{fig:ringGraph}
\end{figure}

For $1\leq j\leq m$, we associate the clause
$C_j=(\ell_{j,1}\vee{}\ell_{j,2}{}\vee{}\ell_{j,3})$ with the set
$O_j$ consisting of three vertices of $R({\cal F})$ representing the
occurrences of the literals of $C_j$ in ${\cal F}$: if
$\ell_{j,1}=x_i$ for some $i\in [n]$ and this is the $r$'th occurrence
of $x_i$ in the clauses, then $O_j$ contains the vertex $v_{i,r}$.  If
$\ell_{j,1}=\bar{x}_i$ for some $i\in [n]$ and this is the $r$'th
occurrence of $\bar{x}_i$ in the clauses, then $O_j$ contains the
vertex $v'_{i,r}$. The other two vertices of $O_j$ are defined
similarly. In our proofs below we will often add a vertex
$c_i$ adjacent to all the vertices of $O_i$ for $i\in [n]$. An example is depicted in
Figure~\ref{fig:ringGraph}.

The following observation which forms the base of many of our proofs
is easy to prove (for a proof of result a very similar to this see
\cite{bangTCS636}).
\begin{theorem}\label{ringgraph}
  Let ${\cal F}$ be a 3-SAT formula and let $R({\cal F})$ be the
  corresponding ring graph.  Then $R=R({\cal F})$ contains a cycle $C$
  which intersects all the sets $O_1,\ldots{},O_m$ so that $R-C$ is a
  cycle $C'$ if and only if $\cal F$ is a `Yes'-instance of 3-SAT.
\end{theorem}

\section{Proof of Theorem \ref{main}}
\subsection{The case $k_1+k_2\leq 3$}

We start with a trivial observation. 
\begin{proposition}
Every graph $G$ with $\delta(G)\geq 1$ and at least 4 vertices has a
($\delta\geq 1, \delta\geq 1$)-partition except if $G$ is a star.
\end{proposition}

\begin{proposition}
  There is a polynomial algorithm for testing whether a graph has a
  ($\delta\geq 1, \delta\geq 2$)-partition
\end{proposition}

\pf We try for every choice of adjacent vertices $u,v$ whether there
is a solution with $u,v\in V_1$.  Clearly $G$ is a 'yes'-instance if
and only if at least one of these $O(n^2)$ attempts will
succeed. Hence by starting with $V_1=\{u,v\}$ and then moving vertices
with at most one neighbour in $V-V_1$ to $V_1$ we either end with a
good partition or $V_1=V$ in which case no partition exists for that
choice $\{u,v\}$.\qed

\subsection{The case $k_1=1$ and $k_2\geq 3$}

The following variant of satisfiability, which we call $\leq 3$-SAT(3),
is known to be {\cal NP}-complete: Given a boolean CNF formula $\cal
F$ consisting of clauses $C_1,C_2,\ldots{},C_m$ over variables
$x_1,x_2,\ldots{},x_n$ such that each clause has 2 or 3 literals, no
variable occurs in more than 3 clauses and no literal appears more
than twice; decide whether $\cal F$ can be satisfied.

As we could not find a proper reference for a proof that $\leq 3$-SAT(3)
is ${\cal NP}$-complete, we give one here as it is presented on pages
281-283 in a set of course
notes\footnote{\url{https://www.csie.ntu.edu.tw/~lyuu/complexity/2008a/20080403.pdf}}
by Prof. Yuh-Dauh Lyuu, National Taiwan University:

Assume that $\cal F$ is an instance of 3-SAT in which the variable $x$
occurs a total of $r\geq 4$ times in the formula (as $x$ or $\bar{x}$)
in clauses $C_{i_1},\ldots{},C_{i_r}$. Introduce new variables
$x_1,\ldots{}, x_r$ and replace the first occurrence of $x$ (in
$C_{i_1}$) by $x_1$ is $x$ is not negated in $C_{i_1}$ and otherwise
replace it by $\bar{x}_1$ in $C_{i_1}$. Similarly we replace the
occurrence of $x$ in $C_{i_j}$, $j\geq 2$ by $x_j$ or
$\bar{x}_j$. Finally we add the new clauses $(\bar{x}_1\vee x_2)\wedge
(\bar{x}_2\vee x_3)\wedge\ldots{}\wedge{}(\bar{x}_r\vee{} x_1)$. These
clauses (which have size 2) will force all the variables
$x_1,\ldots{},x_r$ to take the same value under any satisfying truth
assignment. Repeating this replacement for all variables of the
original formula $\cal F$ we obtain an equivalent instance ${\cal F}'$
of $\leq 3$-SAT(3).

Below we will need another variant which we call $\leq 3$-SAT(5) where
clauses still have size 2 or 3 and each variable is allowed to occur
at most 5 times and at most 3 times as the same literal. By following
the scheme above and for each original variable occurring at least 4
times adding $r$ extra clauses $(x_1\vee \bar{x}_2)\wedge (x_2\vee
\bar{x}_3)\wedge\ldots{}\wedge{}(x_r\vee{} \bar{x}_1)$ we obtain an
equivalent instance ${\cal F}''$ and because the $2r$ new clauses will
form a cycle in the bipartite graph $B({\cal F}'')$ of ${\cal F}''$ it
is easy to see that ${\cal F}''$ is a connected instance of $\leq
3$-SAT(5) if ${\cal F}$ is a connected instance of 3-SAT.  Hence, by Lemma \ref{connected3SAT}, connected $\leq 3$-SAT(5) is ${\cal NP}$-complete.

\begin{theorem}
\label{(1,k)npc}
For all $k\geq 3$ it is ${\cal NP}$-complete to decide whether a graph
has a $(\delta\geq 1,\delta\geq k)$-partition.
\end{theorem}
\pf We show how to reduce an instance of connected $\leq 3$-SAT(5) to
$(\delta\geq 1, \delta\geq k)$-partition where $k\in \{3,4\}$ in
polynomial time and then show how to extend the construction to higher
values of $k$.  We start the construction for $k=3$.

Below we will use several disjoint copies of the graphs $X_{3,1}$ and
$X_{3,2}$ to achieve our construction.  Let $\cal F$ be a connected instance
of $\leq 3$-SAT(5) with clauses $C_1,\ldots{},C_m$ and variables
$x_1,\ldots x_n$. We may assume that each of the $2n$ literals occur
at least once in $\cal F$ (this follows from the fact that we may
assume this for any instance of normal 3-SAT and the reduction above
to $\leq 3$-SAT(5) preserves this property). We will construct
$G=G(\cal F)$ as follows:

  \begin{itemize}
  \item For each variable $x_i$, $i\in [n]$ we introduce three new
    vertices $y_i,v_i,\bar{v_i}$ and two the edges
    $y_iv_i,y_i\bar{v}_i$.
  \item For each $i\in [n]$: if the literal $x_i$ ($\bar{x}_i$) occurs
    precisely once in \cal $\cal F$, then we identify $v_i$
    ($\bar{v}_i$) with the vertex $x$ in a private copy of
    $X_{3,2}$. If $x_i$ ($\bar{x}_i$) occurs precisely twice, then we
    identify $v_i$ ($\bar{v}_i$) with the vertex $x'$ in a private copy
    of $X_{3,1}$.
  \item Now we add new vertices $c_1,\ldots{},c_m$, where $c_i$
    corresponds to the clause $C_i$, $i\in [m]$, and join each $c_j$
    by an edge to those (2 or 3) vertices from
    $\{v_1,\ldots{},v_n,\bar{v}_1,\ldots{},\bar{v}_n\}$ which
    correspond to its literals. If $c_j$ gets only two edges this way,
    we identify it with the vertex $x'$ in a private copy of $X_{3,1}$.
  \item Add $2m$ new vertices $z_1,z_2,\ldots{},z_{2m}$ and the edges
    of the $2m$-cycle $z_1z_2,z_2z_3,\ldots,z_{2m-1}z_{2m},z_{2m}z_1$.
    \item Finally we add, for each $j\in [m]$ the edges $c_jz_{2j-1},c_jz_{2j}$.
  
    \end{itemize}

We claim that $G(\cal F)$ has a $(\delta\geq 1, \delta\geq
3)$-partition if and only if $\cal F$ can be satisfied. Suppose first
that $t$ is a satisfying truth assignment. Then it is easy to check
that $(V_1,V_2)$ is a good 2-partition if we take $V_1$ to be the
union of $\{y_1,\ldots{},y_n\}$ and the $n$ vertices from
$\{v_1,\ldots{},v_n,\bar{v}_1,\ldots{},\bar{v}_n\}$ which corresponds
to the false literals. Note that if a vertex $v_i$ ($\bar{v}_i$) is in
$V_2$ then it will have degree 3 via its private copy of one of the
graphs $X_1,X_2$ or because the corresponding literal occurred 3 times
in ${\cal F}$.\\

Conversely assume that $(V_1,V_2)$ is a $(\delta\geq 1, \delta\geq
3)$-partition. Then we claim that we must have all the vertices
$z_1,z_2,\ldots{},z_{2m}$ in $V_2$: If one of these is in $V_1$, then
they all are as they have degree exactly 3. Clearly we also have
$\{y_1,\ldots{},y_n\}\subset V_1$. However, by construction, the
literal and clause vertices all have degree 3 and they induce a
connected graph (here we use that the instance $\cal F$ has a
connected bipartite graph $B(\cal F)$). Thus all of these vertices
must be in $V_2$, but then each vertex $y_i$ is isolated,
contradiction. Hence all the vertices $z_1,z_2,\ldots{},z_{2m}$ are in
$V_2$ and this implies that all of $c_1,\ldots{},c_m$ are also in
$V_2$. The vertices $y_1,\ldots{},y_n$ are in $V_1$ and hence, for
each $i\in [n]$, at least one of the vertices $v_i,\bar{v}_i$ is also
in $V_1$. Now define a truth assignment a follows. For each $i\in
[n]$: If both $v_i$ and $\bar{v}_i$ are in $V_1$, or $v_i$ is in $V_2$
we put $x_i$ true; otherwise we put $x_i$ false. Since each $c_j$ must
have a neighbour from
$\{v_1,\ldots{},v_n,\bar{v}_1,\ldots{},\bar{v}_n\}$ in $V_2$ this is a
satisfying truth assignment.

To obtain the construction for $k=4$ we replace each copy of $X_{3,1}$
above by a copy of $Y_{4,1}$, each
copy of $X_{3,2}$ by a copy of $X_{4,2}$ and identify each of the literal and clause vertices with the vertex $y$ in an extra private copy of $Y_{4,1}$.
Finally we identify each
vertex $z_t$, $t\in [2m]$ with the vertex $y$ in a private copy of
$Y_{4,1}$.  Now it is easy to see that we can complete the proof as we
did for the case $k=3$.

For all $k\in \{3+2a,4+2a|a\geq 1\}$ we can increase the degree of all
literal, clause and $z_j$ vertices by $2a$ by identifying these with
the $x$ vertices of $a$ private copies of $X_{k,2}$ and repeat the
proof above. \qed

\begin{corollary}
\label{(1,k,k-1)npc}
The $(\delta\geq 1,\delta\geq k)$-partition problem is ${\cal
  NP}$-complete for graphs of minimum degree $k-1$.
\end{corollary}

\pf Recall that in our proof above the vertices corresponding to
clauses must always belong to $V_2$ in any good partition $(V_1,V_2)$
hence if we connect each vertex $y_i$, $i\in [n]$ to
$c_1,\ldots{},c_{k-3}$ by edges we obtain a graph of minimum degree
$k-1$ which has a good partition if and only if $\cal F$ is
satisfiable (The vertices $y_i$, $i\in [n]$ must belong to $V_1$ as
they have degree $k-1$). \qed

\subsection{$(\delta\geq k_1, \delta\geq k_2)$-partition when $2\leq k_1\leq k_2$}

\begin{theorem}
\label{k1,k2bothatleast2}
For every choice of natural numbers $2\leq k_1\leq k_2$ it is ${\cal
  NP}$-complete to decide whether a graph has a $(\delta\geq
k_1,\delta\geq k_2)$-partition.
\end{theorem}

\pf We show how to reduce 3-SAT to $(\delta\geq k_1, \delta\geq
k_2)$-partition. Given an instance $\cal F$ of 3-SAT with clauses
$C_1,C_2,\ldots{},C_m$ and variables $x_1,\ldots{},x_n$ we proceed as
follows. Start from a copy of the ring graph $R=R(\cal F)$ and then
add the following:

\begin{itemize}
\item If $k_2\geq 3$ we identify each vertex of $R$ with the vertex
  $z$ in a private copy of $Z_{k_2}$.
\item For each $i\in [n]$ add a new vertex $u_i$ and join it by edges
  to the vertices $a_{i,1},a_{i,2}$. If $k_1\geq 3$ we identify $u_i$
  with the vertex $z$ in a private copy of $Z_{k_1}$.
\item For each $i\in [n]$ add a new vertex $u'_i$ and join it by edges
  to the vertices $b_{i,1},b_{i,2}$. If $k_1\geq 3$ we identify $u'_i$
  with the vertex $z$ in a private copy of $Z_{k_1}$.
\item For each $j\in [m]$ we add a new vertex $c_j$, identify $c_j$
  with the vertex $z$ in a private copy of $Z_{k_1}$ if $k_1\geq 3$
  and add three edges from $c_j$ to the three vertices of $R$ which
  correspond to the literals of $C_m$.
\item Finally add a new vertex $r$ and join this to all of the
  vertices in
  $\{u_1,\ldots{},u_n,u'_1,\ldots{},u'_n,c_1,\ldots{},c_m\}$ via
  private copies of $W_{k_1}$ by identifying the vertex $u$ with $r$
  and $v$ with the chosen vertex from
  $\{u_1,\ldots{},u_n,u'_1,\ldots{},u'_nc_1,\ldots{},c_m\}$.
\end{itemize}

We claim that the final graph $G$ has a $(\delta\geq k_1, \delta\geq
k_2)$-partition if and only if $\cal F$ is satisfiable. First we make
some observations about $G$:
\begin{itemize}
\item Every vertex $v$ of $R$ which is not a switch vertex has degree
  exactly $k_2+1$ as it has degree 2 in $R(\cal F)$, is adjacent to
  exactly one $c_j$, $j\in [m]$ and if $k_2\geq 3$ then $v$ has been
  identified with one vertex $z$ of a private copy of $Z_{k_2}$.
\item Switch vertices all have degree exactly $k_2+2$.
\item The vertices $u_1,\ldots{}, u_n,u'_1,\ldots{},u'_n$ have degree
  exactly $k_1+1$.
\item All vertices in copies of $Z_{a}$ have degree exactly $a$ when
  $a\geq 3$.
\item All vertices in copies of $W_{k_1}$ have degree exactly $k_1$
\item All vertices $c_j$, $j\in [m]$ have degree $k_1+2$.
\item The vertex $r$ has degree $m+2n$ which we may clearly assume is
  at least $k_1$.
\end{itemize}

For convenience in writing, below we define $Z_2$ to be the empty
graph so that we can talk about $Z_a$'s without having to condition
this on $a$ being at least 3.  Suppose first that $\cal F$ is
satisfiable. By Theorem \ref{ringgraph} this means that $R(\cal F)$
has a cycle $C$ which intersects the neighbourhood of each $c_j$,
$j\in [m]$ and so that $R-C$ is another cycle $C'$. Now we let $V_1$
consist of the vertices of $C$, their corresponding private copies of
$Z_{k_2}$, all the vertices
$\{u_1,\ldots{},u_n,u'_1,\ldots{},u'_n,c_1,\ldots{},c_m\}$ along with
their private copies of $Z_{k_1}$ and finally the vertex $r$ and the
vertices of all copies of $W_{k_1}$ that we used. Let $V_2=V(G)-V_1$,
that is $V_2$ contains the vertices of $C'$ and their private copies
of $Z_{k_2}$. It is easy to check that $\delta(G[V_1])\geq k_1$ and
that $\delta(G[V_2])\geq k_2$ so $(V_1,V_2)$ is a good partition.  Now
assume that $G$ has a good 2-partition $(V_1,V_2)$. The way we
connected $r$ to the vertices in
$\{u_1,\ldots{},u_n,u'_1,\ldots{},u'_n,c_1,\ldots{},c_m\}$ via copies
of $W_{k_1}$ implies that these must belong to the same set $V_i$ as
$r$. If $k_1<k_2$ this must be $V_1$ and otherwise we can rename the
sets so that $i=1$.  Since each $c_j$, $j\in [m]$ has degree $k_1+2$
at least one of the vertices corresponding to a literal of $C_j$ must
belong to $V_1$. Suppose that some vertex corresponding a literal
$\ell$ is in $V_1$, then all the vertices of the path in $R$
corresponding to that literal, including the two end vertices which
are switch vertices, must belong to $V_1$. This follows from the fact
that all these vertices have degree $k_2+1$ and have a neighbour in
$\{u_1,\ldots{},u_n,u'_1,\ldots{},u'_n,c_1,\ldots{},c_m\}\subset
V_1$. Moreover if $a_{i,j}$ (resp. $b_{i,j}$) belongs to $V_2$ then,
as $a_{i,j}$ (resp. $b_{i,j}$) has degree $k_2+2$ and $u_i$
(resp. $u_i'$) belongs to $V_1$, at least one of the vertices of
$\{b_{i,1},b_{i,2}\}$ (resp. $\{a_{i,1},a_{i,2}\}$) belongs to
$V_2$. And as $u_i$ (resp. $u_i'$) belongs to $V_1$, one of
$\{a_{i,1},a_{i,2}\}$ (resp. $\{b_{i,1},b_{i,2}\}$) must lie in
$V_1$. So since $V_2$ is not empty this implies that the restriction
of $V_2$ to $R$ is a cycle consisting of paths $Q_1,\ldots{},Q_n$
where $Q_i$ is either the path $P_{i,1}$ or the path $P_{i,2}$. Hence
$R[V(R)\cap V_1]$ is a cycle intersecting each of the neighbourhoods
of the vertices $c_j$, $j\in [m]$ and hence $\cal F$ is satisfiable by
Theorem \ref{ringgraph}. \qed

\vspace{2mm}

Combining the results of this section concludes the proof of Theorem
\ref{main}.

\section{Higher degrees}

We study the borderline between polynomial and ${\cal NP}$-complete
instances of the partition problems. That is, we try to see how close
we can get to the bound $k_1+k_2+1$ on the minimum degree and still
have an ${\cal NP}$-complete instance. For $k_1=1$ we can give the
precise answer by combining Corollary \ref{(1,k,k-1)npc} and the result
below.

\begin{proposition}
\label{(1,k,k)pol}
There is a polynomial algorithm for checking whether a graph $G$ of
minimum degree at least k has a $(\delta\geq 1, \delta\geq
k)$-partition.
\end{proposition}

\pf It suffices to see that we can test for a given edge $uv$ of $G$
whether there is a $(\delta\geq 1, \delta\geq k)$-partition
$(V_1,V_2)$ with $u,v\in V_1$.  This is done by starting with
$V_1=\{u,v\}$ and then moving vertices from $V-V_1$ to $V_1$ when
these vertices do not have at least $k$ neighbours in $V-V_1$. Note
that this process preserves the invariant that $\delta(G[V_1])\geq
1$. Hence if the process terminates before $V_1=V$ we have found the
desired partition and otherwise we proceed to the next choice for an
edge to start from. \qed

For the $(\delta\geq 2, \delta\geq 2)$-partition problem we can also
give the precise borderline between polynomial and ${\cal
  NP}$-complete instances.

\begin{proposition}
There exists a polynomial algorithm for checking whether a given graph
of minimum degree at least 3 has a $(\delta\geq 2, \delta\geq
2)$-partition.
\end{proposition}
\pf First test whether $G$ has two disjoint cycles $C_1,C_2$. This can be done in polynomial time \cite{bollobas1978,lovaszML16}. If no
such pair exists $G$ is a 'no'-instance, so assume that we found a pair of disjoint cycles
$C_1,C_2$. Now put the vertices of $C_1$ in $V_1$ and continue to move
vertices of $V-V_1-V(C_2)$ to $V_1$ if they have at least two
neighbours in the current $V_1$. When this process stops the remaining
set $V_2=V-V_1$ induces a graph of minimum degree at least 2, since
the vertices we did not move have at most one neighbour in $V_1$. \qed

We now proceed to partitions where $2\leq k_1\leq k_2$ and try to raise the minimum degree above $k_1$ to see whether we can still prove ${\cal NP}$-completeness.

\begin{theorem}
\label{(a,a,a+1)npc}
For every $a\geq 3$ it is ${\cal NP}$-complete to decide whether a
graph of minimum degree $a+1$ has a $(\delta\geq a,\delta\geq
a)$-partition.
\end{theorem}

\pf We give the proof for $a=3$ and then explain how to extend it to larger
$a$.  Let $\cal F$ be an instance of 3-SAT with $n$ variable and $m$
clauses $C_1,\ldots{},C_m$.  Let $R'=R'(\cal F)$ be obtained from
$R(\cal F)$ by adding, for all $i\in [n]$, an edge between all vertices at distance 2 in
one of the paths $P_{i,1},P_{i,2}$, $i\in [n]$ (that is, we replace each of
these paths by their square). Now we construct the graph $H=H(\cal F)$
starting from $R'$ as follows:
\begin{itemize}
\item For each $j\in [m]$: add two vertices $c_{j,1}, c_{j,2}$ and
  join them to the vertices in $R'$ which correspond to the literals
  of $C_j$.
\item add the vertices of a $2m$ -cycle $y_1y_2\ldots{}y_{2m}y_1$.
\item For each $j\in [m]$ add the two edges $y_{2j-1}c_{j,1},
  y_{2j-1}c_{j,2}$ and the two edges $y_{2j}c_{j,2},
  y_{2j},c_{j+1,1}$, where $c_{m+1,1}=c_{1,1}$.
\end{itemize}

The resulting graph $H$ has minimum degree 4 and we claim that $H$ has
a $(\delta\geq 3,\delta\geq 3)$-partition if and only if $\cal F$ is
satisfiable, which we know, by Theorem \ref{ringgraph} and the
previous proofs, where we used the same approach, is if and only if the
vertex set of $R$ (which is the same as that of $R'$) can be
partitioned into two cycles $C,C'$ so that $C$ contains a neighbour of
each of the vertices $c_{j,1},c_{j,2}$, $j\in [m]$.\\

Again the proof is easy when $\cal F$ is satisfiable: Let $C,C'$ be as
above and let $V_2=V(C')$ and $V_1=V(H)-V_2$. It is easy to check that
$\delta(H[V_i])\geq 3$ for $i=1,2$, because $C$ contains a neighbour
of each $c_{j,1},c_{j,2}$, $j\in [m]$ (and we assume that each
literal appears at least twice in $\cal F$ to insure that
$\delta(H[V_2])\geq 3$).  Suppose now that $H$ has a $(\delta\geq
3,\delta\geq 3)$-partition $(V_1,V_2)$. Since adjacent vertices in
$\{y_1,y_2,\ldots{},y_{2m}\}$ have degree 4 and share a neighbour they
must all belong to the same set $V_i$, $i\in [2]$ and this set must
also contain all the vertices $c_{j,1},c_{j,2}$, $j\in [m]$. Without
loss of generality we have $i=1$. Thus $V_2$ is a subset of
$V(R')$. The vertices of $R$ have degree at most 4 in $R'$ and the
initial and terminal vertex of each path $P_{i,1}$ or $P_{i,2}$ has
degree 3. Using this is not difficult to see that if some vertex of a
path $P_{i,1}$ or $P_{i,2}$ is in $V_2$ then all the vertices of that
path and the two adjacent switch vertices are in $V_2$. If there is
some $i\in [n]$ so that both of the vertices $a_{i,1},a_{i,2}$ or both
of the vertices $b_{i,1},b_{i,2}$ are in in $V_2$, then, using the
observation we just made, all vertices of $R'$ would be in $V_2$ which
is impossible. Similarly we can show that $V_1$ cannot contain both of
the vertices $a_{i,1},a_{i,2}$ or both of the vertices
$b_{i,1},b_{i,2}$ for some $i\in [n]$. Hence for each switch
$\{a_{i,1},a_{i,2},b_{i,1},b_{i,2}\}$, exactly one of the vertices
$a_{i,1},a_{i,2}$ and exactly one of the vertices $b_{i,1},b_{i,2}$ is
in $V_2$. Now we see that the vertices in $V_1$ and $V_2$ both induces
a cycle in $R$ and as the vertices $c_{j,1},c_{j,2}$ have degree 2
outside $R$, the cycle in $R$ which is in $V_1$ must contain a
neighbour of each of $c_{j,1},c_{j,2}$, $j\in [m]$. Hence, by Theorem
\ref{ringgraph}, $\cal F$ is satisfiable. \\

We obtain the result for higher values of $a$ by induction where we
just proved the base case $a=3$ above. Assume we have already
constructed $H_a=H_a(\cal F)$ with $\delta(H_a)\geq a+1$ such that
$H_a$ has a $(\delta\geq a,\delta\geq a)$-partition if and only if
$\cal F$ is satisfiable. Construct $H_{a+1}$ from two copies of $H_a$
by joining copies of the same vertex by an edge.  It is easy to check
that $H_{a+1}$ has a $(\delta\geq a+1,\delta\geq a+1)$-partition if
and only if $H_a$ has a $(\delta\geq a,\delta\geq a)$-partition. \qed

\begin{theorem}
\label{(2,3,3)npc}
Deciding whether a graph of minimum degree 3 has a $(\delta\geq
2,\delta\geq 3)$-partition is ${\cal NP}$-complete.
\end{theorem}

\pf Let $X=X({\cal F})$ be the graph we obtain by
starting from the ring graph $R(\cal F)$ and then adding the
following:

\begin{itemize}
\item Add the vertices of a tree $T$ whose internal vertices have all
  degree 3 and which has $2m+4n$ leaves denoted
  by\\ $u_{1,1},u_{1,2}\ldots{},u_{n,1},u_{n,2},u'_{1,1},u'_{1,2},\ldots{},u'_{n,1},u'_{n,2},c_{1,1},c_{1,2},c_{2,1},c_{2,2},\ldots{},c_{m,1},c_{m,2}$,
  where the pairs $u_{i,1},u_{i,2}$ and $u'_{i,1},u'_{i,2}$ have the
  same parent in $T$ for $i\in [n]$ and so do each of the pairs
  $c_{j,1},c_{j,2}$, $j\in [m]$.  Here the vertices $c_{j,1},c_{j,2}$
  correspond to the clause $C_j$
\item Join each vertex $c_{j,1},c_{j,2}$, $j\in [m]$ to the 3 vertices
  in $R$ which correspond to the literals which correspond to $C_j$
  and add the edge $c_{j,1}c_{j,2}$.
\item Add $4n$ new vertices $w_{i,1},w_{i,2},w'_{i,1},w'_{i,2}$, $i\in
  [n]$. Add the edges $w_{i,1}w_{i,2}, w'_{i,1}w'_{i,2}$, $i\in [n]$.
\item For each $i\in [n]$ add the edges
  $w_{i,1}a_{i,1},w_{i,2}a_{i,2},w'_{i,1}b_{i,1},w'_{i,2}b_{i,2}$.
\item For each $i\in [n]$ join each of $u_{i,1}$ and $u_{i,2}$ by
  edges to the vertices $w_{i,1},w_{i,2}$ and add the edge
  $u_{i,1}u_{i,2}$.
\item For each $i\in [n]$ join each of $u'_{i,1}$ and $u'_{i,2}$ by
  edges to the vertices $w'_{i,1},w'_{i,2}$ and add the edge
  $u'_{i,1}u'_{i,2}$.
\end{itemize}

We first prove that in any $(\delta\geq 2,\delta\geq 3)$-partition
$(V_1,V_2)$ of $X$ all the vertices of $T$ must be in the same set
$V_i$. Note that we can not have $c_{j,1}$ and $c_{j,2}$ in different
sets of the partition because then one of their 3 neighbours in $V(R)$
must be in both sets. By a similar argument, for each $i\in [n]$ the
vertices $u_{i,1}$ and $u_{i,2}$ must belong to the same set in the
partition and the vertices $u'_{i,1}$ and $u'_{i,2}$ must belong to
the same set in the partition. It is easy to check that this implies
our claim for $T$.\\

If $\cal F$ is satisfiable, then by Theorem \ref{ringgraph}, we can
find vertex disjoint cycles $C,C'$ in $R$ such that $C$ contains a
vertex corresponding to a literal of $C_j$ for each $j$ and
$V(R)=V(C)\cup V(C')$. Now we let $V_1=V(C')$ and $V_2=V(X)-V_1$. It
is easy to check that this is a $(\delta\geq 2,\delta\geq
3)$-partition because $C$ must contain exactly one of the vertices
$a_{i,1},a_{i,2}$ and exactly one of the vertices $b_{i,1},b_{i,2}$
for each $i\in [n]$.\\ Suppose now that $(V_1,V_2)$ is a good
partition of $X$. By the argument above we have $V(T)\subset V_i$ for
$i=1$ or $i=2$. This must be $i=2$ since in the graph $X-V(T)$ all
vertices except the switch vertices have degree 2. Thus we have
$V(T)\subset V_2$ and each of the vertices $c_{j,1},c_{j,2}$, $j\in
[m]$ have a neighbour in $V(R)$ which is also in $V_2$. As in earlier
proofs it is easy to check that if some vertex of one of the paths
$P_{i,1},P_{i,2}$ is in $V_2$ then all vertices of that path are in
$V_2$.  As in the proof of the previous theorem, we now conclude that
for each switch $\{a_{i,1},a_{i,2},b_{i,1},b_{i,2}\}$, exactly one of
the vertices $a_{i,1},a_{i,2}$ and exactly one of the vertices
$b_{i,1},b_{i,2}$ is in $V_2$. Now we see that the vertices in $V_1$
and $V_2$ both induce a cycle in $R$ and as the vertices
$c_{j,1},c_{j,2}$ have degree 2 outside $R$, the cycle in $R$ which is
in $V_2$ must contain a neighbour of each of $c_{j,1},c_{j,2}$, $j\in
[m]$. Hence, by Theorem \ref{ringgraph}, $\cal F$ is satisfiable. \qed

\begin{corollary}
\label{(a,a+1,a+1)npc}
For every $k\geq 2$ it is ${\cal NP}$-complete to decide if a given
graph with minimum degree at least $k+1$ has a $(\delta\geq
k,\delta\geq k+1)$-partition.
\end{corollary}
\pf This follows by induction on $k$ with Theorem \ref{(2,3,3)npc} as
the base case in the same way as we proved the last part of Theorem
\ref{(a,a,a+1)npc}. \qed
      
\begin{proposition}
\label{(2,3,5)pol}
There is a polynomial algorithm for deciding whether a given graph of
minimum degree 5 has a $(\delta\geq 2,\delta\geq 3)$-partition.
\end{proposition}

\pf Let $G$ have $\delta(G)\geq 5$. By Theorem \ref{kanekothm} and the
algorithmic version of this result from \cite{bazganDAM155} we may
assume that $G$ has a 3-cycle $C$. Denote its vertex set by
$\{a,b,c\}$. If $\delta(G[V-V(C)]\geq 3)$ we are done as we can take
$V_1=V(C)$, so assume there is a vertex $d$ which adjacent to all
vertices of $C$. Then $\{a,b,c,d\}$ induce a $K_4$. If
$\delta(G[V-\{a,b,c,d\})\geq 2$ we can take $V_2=\{a,b,c,d\}$, so we
  can assume that there is a vertex $e$ which is adjacent to all 4
  vertices in $\{a,b,c,d\}$ and now $\{a,b,c,d,e\}$ induce a
  $K_5$. Now if $G[V-\{a,b,c,d,e\}]$ contains a cycle $C'$ we can
  conclude by starting with $V_2=\{a,b,c,d,e\}$ and adding vertices of
  $V-V(C')-\{a,b,c,d,e\}$ to $V_2$ as long as there is one with at
  least 3 neighbours in $V_2$. When the process stops $(V-V_2,V_2)$ is
  a good partition. Hence we can assume that $G[V-\{a,b,c,d,e\}]$ is
  acyclic. If one connected component of $G[V-\{a,b,c,d,e\}]$ is non
  trivial with a spanning tree $T$, then two leaves $u,v$ of $T$ will
  share a neighbour in $\{a,b,c,d,e\}$. Without loss of generality
  this is $e$ and now the $K_4$ induced by $a,b,c,d$ and the cycle
  formed by $e,u,v$ and the path between $u$ and $v$ in $T$ are
  disjoint and we can find a good partition as we did above. Hence if
  we have not found the partition yet we must have that
  $G[V-\{a,b,c,d,e\}]$ is an independent set $I$, all of whose
  vertices are joined to all vertices in $\{a,b,c,d,e\}$. If $|I|\geq
  2$ it is easy to find a good partition consisting of a 3-cycle on
  $a,b$ and one vertex from $I$ as $V_1$ and the remaining vertices as
  $V_2$. Finally if $|I|=1$ there is no solution. \qed
  
\section{Further 2-partition problems}
In \cite{thomassenJGT7} Thomassen proved that every graph $G$ of
connectivity at least $k_1+k_2-1$ and minimum degree at least
$4k_1+4k_2+1$ has a 2-partition $(V_1,V_2)$ so that $G[V_i]$ is
$k_i$-connected for $i=1,2$.

It is natural to ask about the complexity of deciding whether a graph
has a 2-partition $(V_1,V_2)$ with prescribed lower bounds on the
(edge-)connectivity of $G[V_i]$, $i\in [2]$.

We start with a simple observation.
      
\begin{proposition}
There exits a polynomial algorithm for deciding whether a given graph
has a 2-partition $(V_1,V_2)$ such that $G[V_1]$ is connected and
$G[V_2]$ is 2-edge-connected.
\end{proposition}

\pf Suppose first that $G$ is not 2-edge-connected. If $G$ has more
than two connected components it is a 'no'-instance. If it has two
components, it is a 'yes'-instance if and only if one of these is
2-edge-connected. Hence we can assume that $G$ is connected but not
2-edge-connected. Now it is easy to see that there is a good partition
if and only if the block-cutvertex tree of $G$ has a nontrivial block which is a
leaf in the block-cutvertex tree. Thus assume below that $G$ is
2-edge-connected. Now consider an ear-decomposition (sometimes called
a handle-decomposition) of $G$ where we start from an arbitrary cycle
$C$. Let $P$ be the last non-trivial ear that we add and let $u,v$ be
the end vertices of $P$. Then $V_1=V(P)-\{u,v\}$ and $V_2=V-V_1$ is a
good partition.\qed

\2

Perhaps a bit surprisingly, if we require just a bit more for the
connected part, the problem becomes ${\cal NP}$-complete.

Both ${\cal NP}$-completeness proofs below use reductions from a given 3-SAT formula $\cal
F$ so we only describe the necessary modifications of $R({\cal F})$.

\begin{theorem}\label{Ginto2ec,con+cycle}
It is ${\cal NP}$-complete to decide whether an undirected graph $G=(V,E)$ has
a vertex partition $(V_1,V_2)$ so that $\induce{G}{V_1}$ is
2-edge-connected and $\induce{G}{V_2}$ is connected and non-acyclic.
\end{theorem}

\pf We add vertices and edges to $R=R({\cal F})$ as follows:
\begin{itemize}
  
\item For each clause $C_j$, $j\in [m]$ we add a vertex $c_j$ and join
  it by three edges to the three literal vertices of $R$ corresponding
  to $C_j$ (as we did in several proofs above).
\item Add new vertices $c'_1,c'_2,\ldots{},c'_m$ and edges $c_jc'_j$,
  $j\in [m]$.
\item Add new vertices $\alpha_1,\ldots{},\alpha_n$,
  $\alpha{}'_1,\ldots{},\alpha{}'_n$ and the edges
  $\alpha_ia_{i,1},\alpha{}_ia_{i,2},\alpha_i\alpha{}'_i$, $i\in [n]$
  \item Add new vertices $\beta_1,\ldots{},\beta_n$,
    $\beta{}'_1,\ldots{},\beta{}'_n$ and the edges
    $\beta_ib_{i,1},\beta_ib_{i,2},\beta{}_i\beta'_i$, $i\in [n]$.
\end{itemize}

We claim that the resulting graph $G$ has a vertex partition
$(V_1,V_2)$ such that $\induce{G}{V_1}$ is 2-edge-connected and
$\induce{G}{V_2}$ is connected and non-acyclic if and only if ${\cal
  F}$ is satisfiable. Note that, by construction, for every good
partition every vertex of $G$ which is not in $R$ must belong to
$V_2$. In particular if a path $P_{i,j}$ contains a vertex of $V_2$
then all the vertices of $P_{i,j}$ are in $V_2$. Since we want $G[V_2]$
to be connected, the edges between $\alpha_i,\beta_i$, $i\in [n]$ and
$R$ imply that for every $i\in [n]$ at most one of the vertices
$a_{i,1},a_{i,2}$ and a most one of the vertices $b_{i,1},b_{i,2}$ can
belong to $V_1$. This implies that exactly one of $a_{i,1},a_{i,2}$
and exactly one of the vertices $b_{i,1},b_{i,2}$ belong to $V_1$ as
otherwise $V_1$ would be empty. Now it is easy to check that the
desired partition exists if and only if $R$ contains a cycle $C'$ which uses
precisely one of the paths $P_{i,1},P_{i,2}$ for $i\in [n]$ and avoids
at least one literal vertex for every clause of $\cal F$. Thus, by
Theorem \ref{ringgraph}, $\cal F$ is satisfiable if and only if $G$
has a good partition.  \qed

\2

Since we can decide whether a graph has two vertex disjoint cycles in
polynomial time \cite{bollobas1978,lovaszML16} the following result,
whose easy proof we leave to the interested reader, implies that it is
polynomial to decide whether a graph has a 2-partition into two
connected and non-acyclic graphs.

\begin{proposition}
A graph $G$ has a 2-partition $(V_1,V_2)$ such that $G[V_i]$ is
connected and has a cycle for $i=1,2$ if and only if $G$ has a pair of
disjoint cycles and either $G$ is connected or it has exactly two
connected components, each of which contain a cycle.
  \end{proposition}

\begin{theorem}\label{2ecintomindeg2}
It is ${\cal NP}$-complete to decide whether a graph $G=(V,E)$ has
a vertex partition $(V_1,V_2)$ so that each of $\induce{G}{V_1}$ and
$\induce{G}{V_2}$ are 2-edge-connected.
\end{theorem}
\pf Let $\cal F$ be a 3-SAT formula and let $G=G({\cal F})$ be the
graph we constructed in the proof above. Let $G_1$ be the graph
obtained by adding the following vertices and edges to $G$:
\begin{itemize}
\item add new vertices $c''_j$, $j\in [m]$, $q_j$, $j\in \{0\}\cup[m]$ and
  $\gamma$
\item add the edges $c''_jc'_j$, $j\in [m]$
\item add the edges of the path $q_0c''_1q_1c''_2q_2\ldots{}c''_mq_m$
\item complete this path into a cycle $W$ by adding the edges
  $\gamma{}q_0,\gamma{}q_m$
\item add an edge from $\gamma$ to all vertices in
  $\{\alpha{}'_1,\ldots{},\alpha{}'_n,\beta{}'_1,\ldots{},\beta{}'_n\}$.
\end{itemize}

We claim that $G'$ has a vertex-partition into two 2-edge-connected
graphs if and only if $\cal F$ is satisfiable. First observe that in
any good partition $(V_1,V_2)$ we must have all vertices of $W$ inside
$V_1$ or $V_2$. This follows from the fact that each $q_i$ has degree
2 so it needs both its neighbours in the same set. Without loss of
generality, $W$ is a cycle in $V_2$. After deleting the vertices of
$W$ we have exactly the graph $G$ in the proof of Theorem \ref{Ginto2ec,con+cycle} above and it is easy to see that all
vertices not in $V(R)$ must belong to $V_2$ in any good
partition. This implies that $G'$ has the desired vertex-partition if
and only if the graph $G$ has a partition $(V_1,V_2')$ so that
$\induce{G}{V_1}$ is 2-edge-connected and $\induce{G}{V_2'}$ is
connected and non-acyclic. This problem is ${\cal NP}$-complete by Theorem
\ref{Ginto2ec,con+cycle} so the proof is complete.  \qed

\2

By inspecting the proof above it is not difficult to see that the following holds.

\begin{theorem}
  It ${\cal NP}$-complete to decide whether a graph has a 2-partition $(V_1,V_2)$ such that each of the graphs $G[V_i]$, $i=1,2$ are 2-connected.
  \end{theorem}

It may be worth while to try and extend the results of this section to  higher (edge)-connectivities.  

\bibliography{refs}
\end{document}